\documentclass[12pt,preprint]{aastex}  

\slugcomment{Version of \today}
\shortauthors{Smith, Henry \& Vishniac}
\shorttitle{Rotation and cycles in gamma Cas} 
\begin{document} 

\newcommand{\iue}{{\it IUE}}
\newcommand{\rxte}{{\it RXTE}}
\newcommand{\gam}{$\gamma$\,Cas}

\title{Rotational and Cyclical Variability in \gam }

\author{Myron A. Smith,} 
\affil{Catholic University of America,\\
        3700 San Martin Dr.,
        Baltimore, MD 21218;           \\
        msmith@stsci.edu,}

\author{Gregory W. Henry} 
\affil{Center of Excellence in Information Systems,\\
        Tennessee State University,  \\
	3500 John A. Merritt Blvd., Nashville, TN 37209;  \\
        henry@schwab.tsuniv.edu,}

\and
\author{Ethan Vishniac}   
\affil{Department of Physics and Astronomy,\\
        The Johns Hopkins University,  \\
        3400 N. Charles St., \\
	Baltimore, MD 21218;  \\
        ethan@pha.jhu.edu }

 \begin{abstract}

\gam~ is an unusual classical Be star for which the optical-band and hard 
X-ray fluxes vary on a variety of timescales. We report results of a nine-year 
monitoring effort on this star with a robotic ground-based (APT) telescope
in the $B$, $V$ filter system as well as simultaneous observations in 2004
November with this instrument and the {\it Rossi X-ray Timing Explorer 
(RXTE)} telescope.  Our observations 
disclosed no correlated optical response to the rapid X-ray flares in this
star, nor did the star show any sustained flux changes any time during 
two monitored nights in either wavelength regime. 
Optical light curves obtained in the APT program revealed, consistent
with an earlier study by Robinson et al., that 
$\gamma$\,Cas undergoes $\sim$3\%-amplitude cycles with lengths of 60--90 
days. Our observations in 2004 showed a similar optical cycle. Over the
nine days we monitored the star with the {\it RXTE}, the X-ray flux 
varied in phase with its optical cycle and with an amplitude predicted 
from the data in Robinson et al. In general, the amplitude of the $V$ 
magnitude cycles are  30--40\% larger than the corresponding $B$ amplitude, 
suggesting the seat of the cycles is circumstellar.
The cycle lengths constantly change and
can damp or grow on timescales as short as 13 days.  We have also 
discovered a coherent period of 1.21581${\pm 0.00002}$ days in all our 
data, which is consistent only with rotation.  The full amplitude 
of this variation is 0.0060 in both filters, and surprisingly, its 
waveform is almost sawtooth in shape. This variation probably
originates on the star's surface. This circumstance hints at the existence 
of a strong magnetic field with a complex topology and an associated 
heterogeneous surface composition.

\end{abstract}
 

\clearpage

\section{Introduction}
\label{int}

  Discovered as the first of its class in 1867, \gam~ (B0.5\,IV) is by 
definition the prototype of ``classical Be stars."\footnote{We define
classical Be star succinctly as a single, post-ZAMS Be star which has
somehow expelled matter which has settled into a disk confined to the
rotational plane.} Its strong H$\alpha$ emission arises from a disk
that is among a handful that has been observed in the infrared, millimeter,
and centimeter radio regions. Its emission is due in large part to the
disk's high central density of 10$^{13}$ cm$^{-3}$, which is among the highest 
known for classical Be stars (e.g., Waters, Cot\'e, and Lamers 1987).
The disk has been imaged interferometrically in H$\alpha$ flux
by several groups (e.g., Quirrenbach et al. 1996, Tycner et al. 2004)
out to a distance of at least 6R$_{*}$. Measurements of the semiminor
to semimajor axis ratio of this image permit estimates of the rotational 
inclination of the star/disk rotational axis: $i$ = 46$^o$ and 60$^{o}$,
respectively. Berio et al. (1999) have imaged a dense azimuthal
sector of the disk which orbited the star with a period of several years. 
This structure is likely due to a one-armed density pattern that develop 
in perhaps ${\frac 14}$ of all Be stars (Hummel 2000) and are responsible
for the oscillating ratio of the $V$ and $R$ H$\alpha$ emission
components in many of these stars. The effect of the star's radiative wind
and stellar disk on its immediate environment is yet to be determined. 
Recently, Harmanec et al. (2000) and Miroschnichenko et al.  (2002) have 
found that \gam~ is in a 204-day binary with a low eccentricity. 
The evolutionary status of the low-mass secondary is unknown. Given the
orbital separation of this system, it is possible that the gravitational 
perturbations from the secondary are important in truncating the outer 
edge of the disk (Okazaki \& Negueruela 2001).

  Although \gam~ is typical for a Be star in most of these respects, it
is highly unusual in others. Chief among its peculiarities is a rich array 
of variability patterns in the optical, ultraviolet, and X-ray regimes
which extend from timescales of seconds to years
(e.g., Horaguchi et al. 1994, Harmanec 2002).
A proper understanding the interrelationship between these 
variabilities requires a series of dedicated time sequences of 
observations in both wavelength domains. This paper represents a 
continuation of a series of efforts dedicated to determining how these 
various patterns are related using simultaneous or contemporaneous
satellites, including the 
{\it Goddard High Resolution Spectrograph} (GHRS) attached to the {\it 
Hubble Space Telescope (HST)}, the Rossi X-ray Timing Explorer (RXTE) and 
the {\it International Ultraviolet Explorer (IUE).} 

  The coordinated X-ray prong of this campaign consisted of a simultaneous
22 hour monitoring in January of 1996 with the {\it GHRS} and {\it RXTE.}
A quasi-continuum light curve generated from the {\it GHRS} spectra 
exhibited a pair of 1--2\% dips over a few hours. The simultaneous 
{\it RXTE} fluxes showed maxima at these same times (Smith, Robinson, \& 
Corbet 1998). Subsequent analysis showed that the ultraviolet dips cannot 
arise on the Be star's surface and are most likely to occur from 
co-rotating ``clouds" close to the star's surface 
(Smith, Robinson, \& Hatzes 1998, hereafter SRH98). The same
ultraviolet spectra showed many sharp features similar to the long known
blue-to-red ``migrating subfeatures" in the star's optical line 
profiles (Yang, Ninkov, \& Walker 1988, Smith 1995, Smith \& Robinson 
1998). A central conclusion of this program was that the immediate 
circumstellar environment of \gam, even beyond the equatorial plane, is highly 
complex. The system of corotating clouds (some heated and some cooled; 
Smith \& Robinson 2000) alone constitutes a more complex environments 
than is indicated for most, if not all, other Be stars. Moreover, the
clouds require a mechanism to anchor them onto fixed points on the surface.
Such a mechanism may be presumed to be a strong magnetic field, but 
the discovery of a postulated field in this rapidly rotating star seems
beyond the reach of present spectropolarimetric detection devices.
This fact suggests that one must search for indirect indicators of
magnetic field that can test this picture.

  Accordingly, in 1997 Robinson, Smith, \& Henry (2002, hereafter RSH02)
mounted a long-term robotic photometric monitoring campaign on this star in the 
Johnson $B$ and $V$ system using an Automated Photometric Telescope (APT) in 
Arizona. Our program was to search for optical signatures of activity unique 
to \gam~ and especially those correlated with known X-ray activity over a
range of timescales. These included the rotational timescale, estimated
to be near one day, as well as longer-term variations that he {\it RXTE} 
campaigns and earlier observations on other X-ray satellites had suggested
are present.  Soon after they initiated this program, RSH02 identified a
strong pattern characterized as small-amplitude, long
(60--90 days) cycles. The fluxes from 8 {\it RXTE} observations matched these
cycles in length and phase but with amplitudes of nearly 100 times larger.
Because the energy associated with the optical variations is much larger
than in the X-ray variations, this correlation cannot be interpreted simply
as a reprocessing of the X-ray modulation, and one must seek another 
explanation.  One clue to this interpretation is that the optical 
variations have a color consistent with the brightening/color trajectories
that accompany the evolution of the disks of \gam~ and other Be stars.  
This fact suggests that the optical variations arise in the disk. 
Based on both the cyclicity and redness of the optical 
variations, RSH02 conjectured that both the optical variations 
and the generation of anomalous X-rays in $\gamma$\,Cas was produced by 
a Balbus-Hawley instability leading to a disk dynamo. Furthermore, the
winding up of putative field lines from the star with the interaction 
of the keplerian disk would and stretch and sever their connection. The ensuing
reconnection would accelerate disk atoms in the form of high energy beams, 
which would generate X-rays when they impacted the surface of the Be star. 
The observation of absorption systems redshifted to 2,000 
km\,s$^{-1}$ in the {\it GHRS} spectra (Smith \& Robinson 2000) 
is consistent with the geometric and kinematic
requirements of this speculative picture.

  This paper presents the results of searches for 
rapid- and intermediate-timescale correlations of optical and X-ray 
fluxes of \gam.~ In $\S$\ref{cyc} we describe newly found characteristics 
of the long-term cycles found by RSH02 from 9 seasons of optical 
robotic data.  In $\S$\ref{rotper} we report the results of a successful
search from this dataset for a periodicity consistent with the star's 
expected rotational period and discuss its implications for models of 
this star's X-ray emission.

\section{Observations }
\label{obsn}

\subsection{Optical data} 
\label{opt}

   The optical photometry discussed in this paper was acquired with 
the T3 0.4-meter Automated Photometric Telescope (APT) located at
Fairborn Observatory in southern Arizona. The photometer uses a 
temperature-stabilized EMI 9924B photomultiplier tube to acquire data 
successively in the Johnson $B$ and $V$ filter passbands. The observations
of \gam~ were acquired in the following sequence, termed a group observation:
{\it K, sky, C, V, C, V, C, V, C, sky, K}, where $K$ is the check star
HD 5395 ($V$ = 4.62, $B-V$ = 0.96, G8IIIb), $C$ is the comparison star
HD 6210 ($V$ = 5.83, $B-V$ = 0.57, F6~V), and $V$ is the program star
\gam~ ($V$ = 2.15, $B-V$ = $-$0.05, B0.5~IV).  To avoid saturating
the photomultiplier tube when observing \gam, we made the observations
through a 3.8 magnitude neutral density filter.  We used 10 second
integration times for \gam~ and HD~5395 and 20 seconds for HD~6210
and the sky readings.

   Three variable-minus-comparison and two check-minus-comparison 
differential magnitudes in each photometric band were computed for
each group observation and then averaged to create group-mean differential
magnitudes. The group means were corrected for differential extinction
with nightly extinction coefficients, transformed to the Johnson system
with yearly mean transformation coefficients, and treated as single
observations thereafter. Typically, several group observations were made
each clear night at intervals of approximately two hours.  The external 
precision of the group means, based on standard deviations for pairs of 
constant stars, is typically ${\pm 0.004}$ magnitudes  
However, point-to-point observations on intensively monitored nights 
were found to have a mean deviation of only ${\pm 0.003}$ mag. 
Group means with a standard deviation greater than 0.01 mag were discarded.
Further details of telescope operations and data-reduction procedures can 
be found in Henry (1995a,b).

   Our photometric \gam~ observing program began in 1997 September and at 
this writing is still continuing.  Our cutoff date for reporting data in 
this paper is 2006 February.  Typically we succeeded in obtaining a few nights 
of data at the beginning of each observing season in June before the Arizona 
rainy season forced us to close the APT operations for the summer beginning
around 4 July each year.  We resumed the monitoring of \gam~ in mid-September 
each year and continued through the end of each observing season in February.
Our observations during the first observing season between 1997 September 
and 1998 February (JD 2450718 through 2450856) were made using different
neutral density filters for \gam~ and the other stars in the group.
We found it difficult to calibrate the final reduced magnitudes properly.
Therefore, these first-season observations have an undetermined offset with
respect to observations in the rest of the seasons.  The instrumental setup 
was stable for observing seasons two (1998/1999) through nine (2005/2006).  
Our dataset over 9 seasons includes 3157 $B$ and 3135 $V$ observations.
A sample of our differential magnitudes is tabulated 
in Table~1. The full dataset is available in the on-line version of this paper. 
(Entries of "99.999" are placeholders when only a partial observation 
was obtained due to interference by possible faint clouds.)
For plotting purposes, we added to our differential magnitudes the apparent 
$V$ and $B$ magnitudes of the comparison star (m$_V$ = 5.84, m$_B$ = 6.40, 
respectively; after Breger 1974) to establish the proper zeropoint for
our measures of \gam. 

   Beginning in the 2000/2001 season, we made an effort dedicate 
occasional full nights ($\approx$8 hours) to the \gam~ program.
A major effort to observe this star simultaneously with the {\it RXTE}
satellite consisted of our attempting to observe intensively during 
the nights just preceding and following four coordinated nights in 2004 
November, as well as the nights themselves. Altogether, 13 nights of 
the sustained monitorings are included in our dataset, as sell as several
other nights of 5 hours or longer.  The cadence rate of these intensive 
observations was one group cycle every 8 to 4 minutes.

\subsection{{\it RXTE} Data}
\label{rxte04}

  The X-ray component of our program consisted of monitoring $\gamma$\,Cas
with the {\it Rossi X-ray Timing Explorer} ({\it RXTE}) satellite using 
the Proportional Counter Array (PCA), which detects photons in the 
2--30\,keV energy range. We used the FTOOLS reduction package to complete 
the pipeline processing of the data and to generate ``Standard 2" light 
curves with a bin time of 16 seconds.

  This program, designated P90001 in the Guest Observer Cycle 9, was 
designed to monitor this star at the same times our APT system was active
during nighttime in Arizona.  As a hedge against inclement weather,
we asked the {\it RXTE} project to monitor the satellite during 4 orbits
on each of four nights distributed over several days during a time when
$\gamma$\,Cas was situated close to the Continuous Viewing Zone of
the satellite. The project was able to meet this request by allocating
8 hours on each of the nights of 2004 November 5, 9, 13, and 14
(UT dates). Our results met the statistics we expected (i.e., they were
neither lucky nor unlucky) since it turned out that we obtained overlap
with the APT monitoring on 2 of the 4 nights. Counting brief interruptions
from Earth occultation, SAA passages, and a high radiation storm, we 
obtained a total on-target time of 21.3 hours on $\gamma$\,Cas. 

  During nearly all the PCA observations, three of the five PCU 
detectors actively integrated on our target. All count rates represented 
in $\S$\ref{rslt} are scaled by 5/3 to make them directly comparable 
with the rates given by Smith, Robinson, \& Corbet (1998, hereafter SRC98).  
The {\it RXTE}/PCA is an efficient photon detecting system, and the errors 
in the net light curve are elevated by only several percent above the combined 
Poisson errors of the gross and model background fluxes (see also SRC98).

\section{Rapid Optical and X-ray Variability}
\label{rslt}

\subsection{Optical Color Variations}
\label{bvcol}

  The first step in our analysis of the optical APT data was to
determine the mean $\Delta$$B$/$\Delta$$V$ slopes of the variations, 
which RSH02 showed to be dominated by the optical cycles. We plotted the
$V$ against $B$ magnitudes for each of the 9 seasons and 
found that the mean color slope is 0.69. However, we found that these 
slope values seem to cluster around two values of about 0.63 and 0.73. 
For example, there are no season-averaged slopes in the range 0.66--0.70. 
Since the formal significance of each of the seasonal average ratios,
including the observational errors is about $\pm{0.05}$, 
these differences are marginally statistically significant to
3$\sigma$ for the two seasonal group means.
We show an example of these contrasting behaviors for seasons 2000/2001 
(``2000") and 2001/2002 in Figure\,\ref{bvscat}. Inspection of this plot 
discloses that the 2001/2002 magnitudes (squares) have a steeper slope than 
the 2000/2001 data (dots). The respective slopes for these two seasons are 
0.73 and 0.61. 

  The reddish color implied by these ratios implies that the 
cyclical flux changes originate in gas cooler than the effective
temperature of the Be star. Following RSH02, we suggest that this 
region is part of the star's decretion disk for two reasons. First, 
extensive Be disks are well known to contribute to the color of a Be star. 
No other structure (or star) exists near $\gamma$\,Cas
that could be responsible for a 2--3\% variation in the combined 
optical light. Second, this change is consistent
with the color changes observed during the evolution of disks, both in 
other Be stars and $\gamma$\,Cas itself. If continued accumulation of
seasonal averages supports this implied bimodality (4 low values, 5 high), 
it would likely mean that the variations are caused in different spatial 
regions of the disk. However, this speculation will take at least a few 
more years of monitoring to substantiate.

\subsection{Rapid X-ray/optical variations}
\label{rpd}

  Our attempt to obtain comparisons of simultaneous observations of
light and X-ray flux variability was significantly marred by ground or
satellite ``weather" during each of our four nights of observations in 
2004 November. On November 5, we obtained {\it RXTE} data 
during a total span of 3 hours, but about ${\frac 12}$ of this 
collection was interrupted by the detectors' shutting off during the 
{\it RXTE} satellite's passage through the SAA or cloudy weather in
Arizona.  Our most successful 
monitoring lasted 7.2 hours on the night of November 9, though 
it was interrupted by a 2-hour radiation storm. On November 13 
the target was monitored over 1.7 hours, of which about 0.6 hours was
interrupted by a SAA passage. No APT data were collected due to cloudy
weather on November 14. To compare the first three nights of X-ray and 
optical fluxes,
we binned the {\it RXTE} 16-second data to the APT's time sampling of about
10 minutes and compared these means with any APT observations within 
8 minutes of each mean. This gave 4, 24, and 10 paired simultaneous 
observations for November 5, 9, and 13, respectively. The data for
November 5 were insufficient for comparison.

 In Figure\,\ref{xv03} we depict the simultaneous APT and {\it RXTE}
for \gam~ on November 9th. Because the mean was not well determined,
we disregarded the data from November 5 and examined the deviations 
of the X-ray and optical fluxes from their nightly mean values. As the
reader might suspect from visual inspection of Fig.\,\ref{xv03}, the
deviations of the X-ray and APT data show no trend at all.\footnote{We 
obtained the same null result 
when we shifted the X-ray fluxes in time by up to $\pm{14}$ minutes
to account for the contingency that this flux comes from the secondary
of the $\gamma$\,Cas system.} Indeed, the correlation coefficients in
the X-ray/optical fluctuation scatter diagrams are less than ${\pm 0.1}$,
and these values are changed insignificantly when regressions are run
including the effects of errors in the observations. In any case, if
we adopt the ratio of X-ray/optical flux variations from RSH02 and 
1997--2004 seasons in $\S$\ref{cyc} below, namely $\Delta$L$_{x}$/$\Delta$$V$ 
= 3.0/.0366 $\approx$ 80${\pm 20}$, we find that our scatter plots from 
November 9 are significantly different from this relation by 8$\sigma$ 
and 12$\sigma$ for the $B$ and $V$ filters. The same
statement can be made to a significance level of 7$\sigma$ for the
November 13 data. The absence of an optical/X-ray correlation on a
rapid (tens of minutes to a few hours) is consistent with the results 
of SRC98, who found no response in the UV continuum light curve of 1996
March 15 while at the same time {\it RXTE} observed almost continuous 
X-ray flaring.

\subsection{Intermediate timescale optical/X-ray variations}

\subsubsection{Monotonic trends in optical data over several hours}

  In inspecting the nights of intensive APT monitoring, we found
that $\simeq$1\% variations with timescales of up to eight hours 
are common.  Two of these $V$ and $B$ variations, consisting of 
rapid light dips or increases,  were discussed in RSH02 (see their Fig.\,10) 
and were a motivation for the present study. However, we also found times for 
which the optical light brightened or dimmed monotonically during an entire 
8-hour night. Figure\,\ref{mvdlt} exhibits the $V$ and $B$ magnitude
time series for the nights of 2003 November 19 and 20. For both nights
and in both filters the optical light level rises or falls over the course 
of $\approx$8 hours. The significances of these trends are 6.2$\sigma$ and 
4.6$\sigma$ ($V$ filter) and  5.3$\sigma$ and 3.7$\sigma$ ($B$-filter).

It is of interest to examine archival X-ray light curves of
\gam~ to see whether they exhibit a similar behavior. In particular, 
we ask what X-ray changes would be implied by these trends if they
followed the X-ray/optical correlated flux ratio of $\sim$80? 
For the data in Fig.\,\ref{mvdlt}, the optical trends would then translate
to X-ray variations of about 50\%. 
To answer the question of whether the X-ray fluxes show a similar behavior,
one can consult the homogeneous {\it RXTE} light curves published 
in Figure 6 of RSH02.  These data represent the results of six
27-hour monitorings of this star with this instrument during 2000. The
light curves show that over 8 hour stretches one can find five similarly
sustained changes in the X-ray flux having full amplitudes of 50\% or
more. These include three increases during Visits 1, 
2, and 5 of these monitorings and two decreases during Visits 2 and 5.
These variations comprise 5 ``events" in 112 hours of {\it RXTE} observation, 
or one every 22.4 hours. 

 To compare the frequency of optical and {\it RXTE} events, we examined 
the light curves of 13 nights during which the APT monitored \gam~ 
intensively (37--59 observations) in each bandpass. On these nights
we found five nights of sustained increases of about $\simeq$0.01 mags. 
(on HJD - 2450000 = 52230, 52962, 53283, 53319--20), 
two nights of decreases (51866 and 52215), and six nights of 
negligible net change (51831, 51834, 52966--7, 53318, and 53322). In all cases
these trends were significant to at least 3$\sigma$, and the $V$ and $B$
filter data showed the same trends or constancy.
Altogether, the history from the intensive APT observations show seven
up/down trends in 104 hours, or an optical rate of one trend event per 
14.9${\pm 5}$ hours of monitoring. This is comparable to the X-ray rate of 
one event per 22.4${\pm 10}$ hours.  All told, 
on any given night the probability of observing a trend is $\sim$7/13.
We were apparently unlucky not to catch one of these events in any our 
four planned all-night monitorings. In any case, it is {\it plausible,} 
based on these statistical arguments that the two behaviors are related 
to one another, and that the $\Delta$$B$/$\Delta$$V$ ratio during a 
night's observations is close to the ratio found in the correlated 
long-term cycles. The rhetorical question before us is whether 
statistical arguments alone based on nonsimultaneous events are a 
compelling argument for an X-ray/optical correlation.

\subsubsection{How can the intermediate-term X-ray and optical variations 
be understood? }

  Clearly, the above question must be answered in the negative because
frequency arguments based on the present data are insufficient to demonstrate
that the X-ray and optical variations are correlated. However, if this is not
the case, one has the problem of explaining how two new phenomena are produced.
The optical variations, which likewise have a $\Delta$V/$\Delta$B ratio of 
1.20$\pm{.07}$ (e.g., Fig.\,\ref{mvdlt}), are difficult to understand in their 
own right.  The implied color term suggests that they are formed in a cool
(circumstellar) environment. That they are visible at all implies that they
are likely to be formed over volumes considerably larger than the UV-absorbing 
corotating ``clouds" of size 0.2--0.3R$_{*}$. This logic implies that
they are most likely to arise in short-lived dense volumes within the disk.
If future observations continue to imply that X-ray and optical trend-events 
occur together with mutually consistent slopes, it may be possible to 
identify them with local cells that occasionally break away from a generally 
organized global oscillation of the disk. This would explain why these
variations adhere to the X-ray/optical ratio of 80. We note that
such occurrences are actually common in astrophysical dynamos, for example 
manifesting themselves as ``stray" sunspots that appear ``out of phase" 
during solar cycles.

\section{The Long-term Cycles}
\label{cyc}

\subsection{Continued correlation of X-ray and optical cycles }
\label{cycxo}

  A key result of the RSH02 study was the discovery of a correlation
between optical and X-ray cycles of roughly 70 days. This correlation
was made possible by a total of eight 27-hour {\it RXTE} observations
that were contemporaneous with the APT observations in 1998 and 2000.
In this paper we report on {\it RXTE} observations that span a
total of nine days. Although this is a much shorter time than a cycle
length, the phasing of the new {\it RXTE} data near the maximum of the 
optical cycle allow one to predict an X-ray flux at the peak of its 
cycle and to compare it with the observations.
Moreover, our range of nine days over which the {\it RXTE} observations 
were made in 2004 November is large enough to compare with the {\it slope} 
of the predicted X-ray variation inferred from the ephemeris for the 
optical cycle during the 2004/5 season, as discussed in the next section. 
The maximum for the sinusoid is taken from X-ray flux maximum (90 
cts\,s$^{-1}$) that correspond to the optical maximum, according to RSH02.
Fig.\,\ref{xrlcnov} exhibits this comparison. The
dots represent the individual X-ray observations. The scatter in 
these data are mainly fluctuations in the X-ray fluxes caused by flares
and changes in the basal flux contribution. The dashed line in this
figure is the RSH02 prediction for the maximum of an optical cycle. 
We emphasize that other than applying the X-ray/optical conversion factor 
of 80, {\em no scaling or adjustments have been made in constructing the
sinusoidal curve in this figure.} The agreement of the trend represented
by the dashed curve to the data represents a confirmation of the
correlation of the X-ray and optical cycles found by RSH02.

\subsection{The optical cycles}
\label{cyco}

  In this section we consolidate the data for nine seasons of APT 
observations of the cycles of \gam.~  For each season we have fit the 
data to a suitably modified sine curve. We started from the results of 
RSH02 that the mean light level, semi-amplitude, period may change during 
a season. In our graphical solutions we have permitted these parameters 
to float freely, but we left the time of zero phase fixed. We modified
the period by introducing a fixed rate of change through
a season, that is, by assuming that \.P is a constant. The
modified representation becomes: 

\begin{equation}
m_v = m_o(t) + a(t)\,sin(\,[2\pi/{\dot P}]\,\,ln( 1 + [{\dot P}/P_o]\,\,t ) + 
\phi_o\,)   
\end{equation}

\noindent where a(t) and m$_{o}$(t) are linear representations in time. 
The logarithm in equation (1) supplies higher order terms beyond the
first-order expansion term in ${\dot P}$ used by RSH02.  This fact 
explains slight differences we have rederived for the 1999/2000 and
2000/2001 cycles. However, like RSH02, we find that the periods derived
for these seasons are too imprecise to link the cycles of these two 
seasons by a simple linear interpolation. RSH02 also noticed
that they were unable to interpolate linearly between the seasonal periods
of 65 and 79 days for 1999 and 2000, respectively, without postulating
phase changes between these seasons. Likewise,
in the current analysis we have found other cases where a period
generated from the data within a season fits the extrapolation from the
previous season. In some cases these mismatches cannot be easily accommodated
by assuming a smoothly increasing or decreasing period between the seasons.
Because this may actually the rule, it is more appropriate to examine the 
behavior of the cycles within each observing season rather than potentially 
overlooking interesting physics implied by these anomalies.

  Figure\,\ref{allcyc} shows our best modified sine curve
fits to the 1997--2004 cycle data for the $V$ filter. Results for the 
$B$ filter are virtually the same, except that the amplitudes are 
slightly smaller. We note also that these data are uncorrected for the
small-amplitude, 1.2-day period discussed in the next section.
Table\,2 gives the corresponding parameters for each 
season, including the start and end mean magnitudes in our solution. 
All values are given in days (HJD) and magnitudes. Errors have been computed
by investigating the values needed to give just unacceptable fits for each
season and taking averages of the results. These errors have meaning only
in the context of our models for individual seasons.
The reader should note that a negative value for the dimensionless
quantity $\dot{P}$/$P$ corresponds to a lengthening period.
The mean full amplitude for the cycles in the $V$ filter, computed as
twice the sum in quadratures of the two semiamplitudes given in the table,
is 0.0366 magnitudes. These amplitudes and the X-ray amplitude in 
Fig.\,\ref{xrlcnov} account for the $\Delta$L$_{x}$/$\Delta$$V$ of 80 we 
have adopted. In constructing these fits we noticed the following departures 
from constant-amplitude fits: 

\begin{itemize}

\item During at least two seasons, 1998/1999 and 2003/2004, the cycles are
exponentially damped. Their damping constants are 50 days and 13 days, 
respectively.  Surprisingly, the 2003/2004
light curve grows again exponentially to an even larger amplitude than it
had at the beginning of the season. The damping and growth rates are
both 13 days. It is further remarkable that little or no phase change occurs
during this transition. 

\item The damping behavior could mislead one to interpret the lengths of
some of the cycles to be about double their true values. Our discovery
that the cycles can damp out, particularly near an extremum, suggests that the 
cycle lengths in \gam~ occur over a slightly more restricted range of about 
60--90 days than recognized by RSH02. For example, the assignment of 55 days
for the period of the 1998/9 was the unforeseen result of not appreciating
the damping behavior.

\item It is also likely (see Fig.\,\ref{allcyc}) that the 1999/2000 
season is affected by a similar, though longer, growth in amplitude
as compared to the previous season. RSH02 found that cycle amplitudes can 
change from one year to the next. We see now that the timescale over 
which this occurs can vary significantly.

\item Changes in the period can occur over intervals shorter than 
the period itself (RSH02's ``change in phase").  Although in  
Fig.\,\ref{allcyc}, we have represented the 1999-2000 cycles season by 
both a simple (but growing in amplitude) sine wave and one with an 
increasing period (dashed line), it is also possible to fit the data 
with a sine curve in which the phase seems to lag by 20--30 days
at the minimum of the second cycle in this season.

\item The periods can change in more complex ways than as represented by 
a single ${\dot P}$ term in equation (1). For example, the data in the 
2002/2003 season can be fit adequately only by allowing the ``period" 
to shorten and then lengthen during this interval. Thus, neither the 
dashed nor solid-line representation, denoting a constant period or 
constant ${\dot P}$, respectively, can represent the obvious change 
in intervals between the three minima of this season.

\item The mean magnitudes change slowly from season to season, for 
example undergoing variations from 2.171 to 2.188 magnitudes from
1998/9 to 1999/2000. These changes, together with growth/dampings
of cycles, can produce ambiguities in the interpretation
of the cycle length. For example, we are unable to model the variations
during the 2005/6 season yet because we do not know how to interpret
the mean magnitude for this season. The light curve at this time 
seems to undergo part of a long cycle but then to settle down to a
short cycle length ($\approx$50 days). Future monitoring may help resolve
the evolution of the light curve when at least three cycle parameters
change during a season.

\end{itemize}

 An examination of these fitting parameters indicates no correlation 
among them, for example of amplitude with time, nor are they correlated
with the yearly $\Delta$$B$/$\Delta$$V$ values or with the much longer
$V/R$ H$\alpha$ emission data given by Miroschnichenko et al. (2002). It is
much too early to be able to associate cycle attributes with one another.

 Finally, we point out that RSH02 found the correlated optical/X-ray 
cycles from the 1999 and 2000 cycles in this plot. In addition, we note 
that the slightly low X-ray flux recorded during November 1998 
(Robinson \& Smith 2000) corresponds to the weak secondary minimum of 
the 1998/9 optical cycle shown in Fig.\ref{allcyc}.

\subsection{The Viability of the Dynamo Model}

The suggestion in RSH02 that the correlated variations in X-rays and optical
emission are ultimately due to dynamo cycles in the decretion disk remains an
attractive hypothesis. However, there are significant differences between the
dynamics of this system and any of the current numerical simulations of disk
dynamos in the literature. 
The bulk of the decretion disk mass, and most of its optical emission, 
comes from the inner regions of the disk, where the magnetic field 
of the star may be strong enough to enforce corotation out to at least
$\sim 1$$R_{*}$ above the star's surface (RSH02). If one takes the  
characteristic density of the disk to be 10$^{12-13}$ cm$^{-3}$, then the 
strength of the stellar magnetic field in pressure equilibrium with respect 
to the bulk kinetic (i.e., orbital) energy should be a few hundred 
Gauss near the corotation radius.
This is consistent with a typical magnetic field strength at the stellar 
surface of more than $10^{4}$ G (see \S 5.4) in a highly disordered field.
On the other hand, a disk dynamo will be driven by the magneto-rotational
instability (MRI), which will saturate at a magnetic energy density less 
than the thermal pressure in the disk. This density corresponds to a magnetic 
field strength of $\sim 10$ G.  Numerical simulations (for a review see Balbus 
\& Hawley 1998) suggest that magnetic fields may fail to reach this limit 
by an order of magnitude, depending on the orientation of the field  
in the disk.  Apparently, in the absence of mitigating factors,
we might expect the stellar magnetic field to dominate over any plausible 
MRI dynamo field by a factor of roughly 100.  Such a strong externally 
driven field would act to suppress the local MRI instability, although 
the global stability of the system is uncertain  (Spruit, Stehle \& 
Papaloizou 1995; Stehle \& Spruit 2001).  We conclude that unless arguments 
can be advanced that the stellar field is somehow excluded from the disk, for 
example by the stellar wind (see, e.g., ud Doula \& Owocki 2000), its 
influence on the disk field could quench the MRI dynamo mechanism.

  Despite this problem, it is still possible that long timescale variations 
in $\gamma$\,Cas are the result of disk instabilities.  One possibility 
is that the dynamo mechanism may have little to do with the MRI dynamo.
Instead, they may be a heretofore uncontemplated result of nonlocal 
interactions between the disk and the star.  
In either case it is clear that the available simulations of the 
disk dynamos are an unreliable guide to the dynamics of this system.

\section{Discovery of a 1.2-day periodicity }
\label{rotper}

\subsection{Observational characteristics of a 1.2-day periodicity }
\label{rotcoh}

The initial motivation to monitor \gam~ with the APT was to search for a 
signature of the rotational period.  For a broad-lined, early B-type star 
on the main sequence, this period should be near one day.  To search for
such a period, we analyzed the nine individual seasons of our \gam~ 
observations with the method of Vani\^cek (1971). This procedure is based
on least-squares fitting of sinusoids.  Henry et al. (2001) and additional
references therein describe how this method allows us to locate and fix 
individual frequencies in succession to determine all of the multiple 
frequency components present in a dataset.  The Vani\^cek (1971) method 
differs from prewhitening for a given frequency before searching for the 
next by the investigator's fixing only the given frequency and not its 
amplitude, phase, or mean light level before computing a new power spectrum.  
The new frequency search is carried out while simultaneously fitting a single
new mean brightness level along with the amplitudes and phases of all
frequencies introduced as fixed parameters. In the resulting least-squares
spectra, we plot the fractional reduction of the variance (reduction
factor) versus trial frequency.  This method holds an advantage over 
prewhitening, especially in the low-frequency domain, where mean light 
levels, amplitudes, and phases might be poorly determined. Thus, it avoids 
perpetuating systematic errors in these parameters in successive searches for 
additional frequencies.

We applied this strategy to each of the nine seasons of our \gam~ photometry
by searching the frequency range 0.001 to 2.5 day$^{-1}$, corresponding
to a period range of 0.4 to 1000 days.  In each season, for both the $V$
and $B$ datasets, we found evidence for a weak frequency near 
0.8225 day$^{-1}$, or approximately 1.215 days, after fixing up to several 
low frequencies arising from the cycles discussed in \S 3.4 above.  Due to 
the changing nature of the stronger cycles and the difference in the cadence 
of the observations from season-to-season (\S 2.1, above), the robustness 
of the 1.215 period we determined varied significantly from year to year.

As an additional check on the presence of the 1.215 day period, we repeated 
our period analysis on the combined season 2 through 9 datasets; season 1
was omitted from this analysis because of the undetermined offset of those 
observations (see \S 2.1, above).  To our surprise, the 0.8225 day$^{-1}$
frequency was clearly seen in the power spectrum of the complete dataset
after fixing several of the strongest low frequencies, suggesting the
possibility that the 1.215 day period remains {\it coherent} in phase 
throughout our entire 9 year dataset.  The top panel of Fig.\,\ref{pspect} 
shows the power spectrum of the season 2 through 9 $V$-filter observations 
after fixing the low frequencies 0.00838, 0.01225, 0.00103, and 
0.00321 day$^{-1}$.  The strongest remaining frequency, marked with the 
large arrow, is 0.82250 $\pm$0.00001 day$^{-1}$, corresponding to a period of 
1.21580 $\pm$0.00002 days.  The smaller arrows mark the $\pm$1 day aliases of 
the 0.82250 day$^{-1}$ frequency.  The small clusters of frequencies at 1 and
2 day$^{-1}$ are the one day aliases of the residual low-frequency 
variations visible at the extreme left of the top panel that have not been
fixed in this analysis.  The bottom panel of Fig.\,\ref{pspect} shows the 
results of fixing the previous four low frequencies {\it and} the 0.82250 
day$^{-1}$ frequency.  There are no other frequencies in this range that 
appear significantly above the noise level.  We repeated this analysis with 
the season 2 through 9 $B$-filter observations with nearly identical results. 
After fixing three low frequencies at 0.00837, 0.01222, 0.00383 day$^{-1}$ 
in the $B$ data, the strongest remaining frequency was 0.82249 $\pm$0.00001 
day$^{-1}$, corresponding to a period of 1.21582 $\pm$0.00002 days.  Thus, 
the mean period from the $V$ and $B$ datasets is 1.21581 $\pm$0.00001 days, 
and it is stable against removal of the low frequencies in both cases. An
additional search for higher frequencies out to 30 day$^{-1}$ was negative. 

 The top panel of Figure\,\ref{phsv29} plots the variations of all the 
$V$ observations from seasons 2 through 9 after 
prewhitening for the four low frequencies mentioned in the above paragraph
and phased with the 1.21581 day period and the arbitrary epoch HJD\,2450000.
A similar plot for the $B$ observations is essentially identical. The 
peak-to-peak amplitudes of the 1.21581 day period in the $V$ and $B$ 
passbands, based on least-squares sine fits of the non-prewhitened original 
datasets, were 0.0053 $\pm$0.0004 and 0.0055 $\pm$0.0004 mag, 
respectively.  Thus, the amplitudes are identical within their uncertainties.

Inspection of Fig.\,\ref{phsv29} reveals that the waveform is highly 
nonsinusoidal.  Indeed, the maximum occurs at phase $\sim$0.49, while the 
minimum is visible at about 0.71.  Such an asymmetric waveform is unusual in 
astrophysical processes that generate a single oscillation.  This property 
suggested that the waveform could be fit analytically by a Lehmann-Filhes 
solution, in the manner of fitting radial velocity solutions of a binary 
star. Then the functional form of the single-wave curve is given by:

\begin{equation}
m\,\,=\,\,K\,cos(\phi + \omega) + e\,cos( \omega).  
\end{equation}

\noindent Here $m$ represents the apparent magnitude rather than a radial 
velocity, and $K$, $\phi$, $e,$ and $\omega$ can be thought of as analogs 
of the orbital parameters velocity semi-amplitude, phase, eccentricity, 
and longitude of periastron.  The latter quantity is an arbitrary phase 
zeropoint similar to our arbitrary photometric epoch of JD 2450000.  We 
verified the aptness of this functional fit by binning the phase curves into
100 bins (0.01 cycles) and overplotting the 0.01-cycle means with the solution.
This is shown in the bottom panel of Fig.\,\ref{phsv29}.
No systematic departures of the binned data could be discerned with respect
to the solution in either dataset, further suggesting that the 1.21581 day
period is coherent throughout our dataset.  For the $B$ filter our solution 
is K = 0.00300${\pm .00013}$ mags, $e$ = 0.333${\pm 0.049},$ and $\omega$ = 
280.3${\pm 9.9}$, while for the $V$ filter it is K = 0.00302${\pm .00013}$ mag, 
$e$ = 0.372${\pm 0.052},$ and $\omega$ = 289.6${\pm 9.7}$.  
Since this waveform is to be preferred over the sinusoidal representation, 
the derived $K$ values should be regarded as the true semi-amplitudes of the
variation. We also note the remarkable fact that the $K_{B}$/$K_{V}$ ratio 
is 1.00${\pm 0.007}$. 

As a final check on the long-term amplitude and phase coherency of the 
1.21581 day period, we fit least-squares sinusoids at that period to the 
non-prewhitened data in the individual observing seasons 2 through 9
(1998/9--2005) independently.
Table\,3 lists the resulting peak-to-peak amplitudes in $V$ and phases of 
minimum for the eight observing seasons in columns 2 and 3, respectively, 
along with their individual formal errors.  Notice that
standard deviations of the 
amplitudes and phases are 0.0030 mag and 0.074 phase units, respectively, 
both significantly larger than the typical formal errors.  This could suggest
conceivably that there are season-to-season variations in amplitude and 
phase and, thus, some degree of non-coherency in the 1.21581 day period.  
However, as noted 
earlier, the natures of the cycles and the cadences of the observations 
are significantly different from season to season.  This situation could 
result in systematic errors in the amplitudes and phases that would render 
the formal errors too small.  To test this hypothesis, we removed 
(prewhitened) the 1.21581 day period from the season 2 through 9 $V$ 
light curves and added a similar but randomly chosen 1.1976 day coherent 
variation with the same amplitude as the 1.21581 day period.  We then 
repeated the sine curve fits for the individual seasons using the artificial 
1.1976 day period.  The results are given in columns 4 and 5 of Table~3.  
The standard deviations of the amplitudes and phases of the artificial 
1.1976 day period are 0.0025 mag and 0.089 phase units, similar to our 
results with the 1.21581 day period and also larger than the typical formal 
errors.  Therefore, this result is consistent with the amplitude and phase 
stability of the 1.21581 day period throughout our complete dataset.
This addresses the disparity between the ``sigma'' in the table
and the formal photometric errors.

\subsection{The rotational nature of the 1.2-day period}
\label{rot12}

  Three possible explanations present themselves for a coherent 1.2 day
period in an early B star:
rotation, ellipsoidal variation, and (nonradial) pulsation.
Among these, rotation is by far the most likely mechanism, primarily
because it is well within the narrow range of rotational periods 
suggested for this star. The rotational period can be determined by
knowing the rotational velocity, radius, and rotational obliquity, $i$.
Recent values of the rotational velocity are 400 km\,s$^{-1}$ and 
432 km\,s$^{-1}$ (Harmanec 2002,  Zorec, Fr\'emat, \& Cidale 2005), the
mass and log gravity are 15$.2{\pm 0.9}$M$_{\odot}$ and 3.80, respectively.
As already mentioned, values of $i$ ranging from 46$^{o}$ to 60$^{o}$
have been determined from interferometry. This range of parameters 
gives an expected range for the rotational period of 1.08--1.41 days. 
The mean of these is 1.245 days, which is close to the 1.21581 period
from our APT photometry.  However, before accepting rotation as the
driver of this variation, we consider the competitive explanations.

  One may dismiss readily the possibility that the 1.2 day
period arises from an ellipsoidal variation of the Be star in a 2.4
day orbit. Such a variation could only arise in a binary with comparable
component masses in a close system. It would produce a radial velocity 
variation, 2$K$, of tens of km\,s$^{-1}$, which would have 
been easily observed, e.g., in the data of Harmanec et al. (2000) and
Miroschnichenko et al. (2002). In addition, a 2.4-day system would 
be tidally locked for a main sequence B star. The photospheric line
profiles are much too broadened to arise from a star having this
long a rotational period.
 
  Arguments against nonradial pulsations are also strong, but not as 
clear cut. Empirically, we know that NRPs in B stars are often multiperiodic.
This circumstance has opened up the new field of asteroseismology in 
massive stars. However, the best examples of pulsations in massive
stars hotter than the short-period $\beta$\,Cep stars, namely $\zeta$\,Oph 
and HD\,93521 (Reid et al. 1993, Howarth \& Reid 1993), indicate that 
the pulsations in these stars have periods of only several hours. 
A stronger, theoretical argument is that the
NRP periods of B stars are {\it predicted} to decrease with T$_{\rm eff}$ 
because their values should be of the order of the thermal timescale in the
Z-bump driving zone where the pulsations are excited (e.g., Pamyatnykh 
1999). These timescales for $\sim$B0.5\,IV stars are only several hours 
and therefore appear incompatible with the 1.2 day period. 

  Additional (in our view, clenching) arguments against NRP are the
complete absence of color variation in our well constrained determination 
of the $\Delta$$B$/$\Delta$$V$ ratio and the nearly saw-tooth waveform. 
To our knowledge this flux ratio is grayer than any known variation
among BAF-type pulsating stars. 
In contrast, variations imposed by low-frequency (co-rotating frame!) 
NRP modes in B stars show a color variation (e.g., De Cat et al. 2005), 
and this is consistent with theoretical predictions that a color term
should be generated in the flux variations. In addition, the waveform 
exhibited by NRP variables across the H-R Diagram is universally close 
to sinusoidal.

\subsection{Previous claims of a rotational period} 

  The present period is at variance with several claims of a rotational
period for \gam~ in the literature. These include claims by Harmanec (1999) 
of a spectroscopic period. More perplexing are reports by Marchenko et al. 
(1998) and Harmanec et al. (2000) of a total of 3 periods in the range of
1.04--1.64 days in the {\it Hipparcos} (Perryman et al. 1997) light curve. 
In addition, SRC98 and SRH98 found a period of 1.123 days from {\it RXTE} 
and {\it IUE} fluxes. We discuss each of these as follows.

   Harmanec (1999)'s estimate of 1.16 days, close to our own, was based on
the assumption that the acceleration of msf's in optical line profiles 
is due to fixed disturbances
on the star's surface. This value, unlike ours, is subject to uncertainties
in the estimated inclination and radius of the star, and especially to the
distance of the clouds responsible for msf's above the star's surface.
placement of the msf disturbances on the star's surface. His result can be
expected to be imprecise, but rather close to our own, as indeed it is.

 We have repeated the analysis of Marchenko et al. (1998) Harmanec et al. 
(2000). We agree with the latter authors that the {\it best candidate 
period} in the  {\it Hipparcos} light curve is 1.487 days, However, when
the dataset is broken into two and three segments, this signal was found to 
be confined to the middle segment and therefore cannot be described as a
coherent period. From our examination of this light curve, we conclude that 
no coherent periods near 1 day can be reliably found. The principal problem 
is that the {\it Hipparcos} data are too sparse and poorly sampled to resolve 
the long-term, low-amplitude (3\%) 60--90 day irregular cycles in \gam.
It appears that the short periods derived by these authors are likely to be 
aliases of these cycles.  Our conclusion is confirmed by our experience with 
comparisons of other extended APT datasets and corresponding {\it Hipparcos} 
light curves. For example, Henry et al. (2000) have used comparable quality 
APT data for a large sample of stars, many of which were found to be 
variable with full amplitudes as low as 0.6\%. These same stars were also 
observed by {\it Hipparcos.} Variability was generally not found in the
latter light curves when the amplitudes were less than 3\%.  Similarly, 
Henry \& Fekel (2002a, 2002b) used APT data to discover periods in 6 new 
$\gamma$\,Dor and 5 $\delta$\,Scu variables, respectively.  Only two of 
these stars showed indications of variability in the {\it Hipparcos} data.
However, because two-channel photometers were used, this probably slightly 
overstates the relative APT advantage.

  The 1.21 day period likewise contradicts the SRH98 claim of a period near 
1.12 days.  This claim was based from bootstrapping from an initial rough 
(and nonunique) estimate obtained from an analysis of UV continuum dips in 
a 1.2 day time series of {\it IUE} observations.  SRC98's supposed that
these dips correspond to repeating X-ray maxima over many 
rotations.  Based on their search for recurring marker in a series of 6 
1.1-day {\it RXTE} observations, RSH02 concluded that such markers may 
disappear within a week, or about 6 rotation cycles. Indeed, this 
discovery undermined a key assumption made in deriving the SRH02 period.

\subsection{The origin of the periodicity}
\label{atm12}

  Explanations for the origin of this period are not yet well constrained,
but there is perhaps enough information to point us in the right 
direction. Possible locations of the origin of the rotational signature
are the surface of the star and a co-rotating structure just above some point 
above the stellar surface. The sawtooth waveform presents a difficulty 
for either case. The circumstellar explanation at first seems attractive 
because we know that such clouds exist. The absence of a color variation
means that the continuum-emitting clouds would have to have the photospheric
temperature and thus be close to the 
star. In this picture, the fact that the light level falls below the 
maximum during most of the cycle would mean that emitting clouds are
distributed over a range of stellar longitudes and are occulted as they
co-rotate behind the star. This requirement for proximity to the surface
forces the circumstellar explanation to posit the existence of many small 
dense sources corotating over points on the surface with a range of 
longitudes. These sources would also have to be optically thick in the 
continuum, and this requires a column density of at least 10$^{25}$ 
cm$^{-2}$. This requirement exceeds by two or three orders of magnitude 
the thick component of a two-density model of small circumstellar clouds
discussed by Smith \& Robinson (1999), based on their 
analysis of the variable Si\,IV and S\,IV
line absorptions. While this is not impossible, there is no support for
continuum reemission from the cloud properties, for example, from the
flatness of the 1996 March {\it GHRS} light curve between its obvious dips.
These arguments cast doubt on a circumstellar origin for the variations
in the optical light curve.

   The explanation we favor is one in which a structure is firmly 
rooted within the star's envelope and is visible to observers at the 
surface. To date, we are aware of only one other early-to-mid B-type 
star, HD\,37776, that exhibits rotational modulations (period = 1.538675
days; Adelman 1997) in unpolarized optical-continuum 
light.\footnote{By this statement we are excepting the case
of the B2p star $\sigma$\,Ori\,E, which also exhibits continuum light 
variations (Hesser, Moreno, \& Ugarte 1977). The difference is that its 
photometric variations are the result of absorption from an intervening 
corotating cloud (Smith \& Groote 2001). In contrast, the photometric
variations in HD\,37776 are well correlated with strengths of spectral lines, 
also arising from high excitation potentials, e.g., by Khokhlova et al. 
(2000), and these cannot arise in an unheated circumstellar cloud.}
This star is special even among Bp stars cause of its especially strong
dipolar field strength (60\,kG). The dipole coexists with a quadrupolar
component that is nearly antialigned with the primary dipole (Thompson \&
Landstreet 1985, Khokhlova et al. 2000). The ensemble produces a double
wave light curve. In the visible/red bandpasses the presence of the secondary
flux bump is suppressed, with the result that the light curve takes on
a quasi-sawtooth waveform (Mikulasek et al. 2006). The star has a
heterogeneous surface distribution of the metals, and this appears to
be the source of the light curve variations (Khokhlova et al. 2000).

  The example of HD\,37776 suggests that large and multi-polar magnetic 
fields on an early-type B star can produce its observed photometric
variations over the rotational period. 
The amplitude of the visible wavelength light curve is about 3.3$\times$
larger than the variations we have reported for $\gamma$\,Cas. A simplistic 
scaling of the photometric amplitude ratio of HD\,37776 
with respect to $\gamma$\,Cas would likely lead to an overestimate of
its mean surface field strength. Indeed, the details of the magnetic 
topology dominate such estimates and preclude a ready estimate of the 
maximum field strength. 

  In view of these arguments, we believe the best explanation for the 
monochromatic variations of \gam~ is that they are produced by an
undiscovered strong multipolar field rooted on magnetic poles that are
distributed over a range of stellar longitudes. (We emphasize once
again that due to the broad lines of this star's spectrum, complicated
by emission, an actual field is not likely to be detected for some
time). The requirement that the fields be multipolar and distributed
across the surface, is supported by the broad distribution of X-ray active
maxima occurring over several 27-hour long X-ray monitoring campaigns 
(essentially the rotational period) of \gam~ (e.g., RSH02). 
Indeed, the instances of high X-ray activity are so numerous that it is
possible only to discover {\it inactivity} markers by cross-correlating
reciprocal X-ray flux curves (Robinson \& Smith 2002, RSH02). Finally,
the absence of a single dominant dipolar field is suggested by the lack
of evidence of a magnetically focused wind, i.e., modulated low-velocity 
emissions and absorptions of UV resonance lines (e.g., Shore \& Brown 1990), 
In contrast, these variations are the rule among well-observed magnetic 
Bp stars.

\section{Conclusions}
\label{concl}

  However well studied, \gam~ has become a prototypical astronomical
onion, with each new discovery raising many more questions than answers
about the interaction of complex processes in hot unevolved stars. In broad
strokes, we can summarize the phenomenology relating to the complicated
circumstellar environment by the following description.

  A number of recent studies of \gam, including the simultaneous {\it
RXTE}--{\it GHRS} campaign of 1996, have shown that in addition to the
continuum flux, the strengths of a number of UV absorption lines are strongly
correlated (or anti-correlated) with X-ray flux. Altogether, there are
at least three systems of circumstellar debris: the (mainly)
keplerian orbiting disk, the co-rotating clouds of various temperatures
(visible in the UV continua or in the UV and optical as migrating subfeatures),
and redshifted blobs moving at least roughly toward the star.
There is therefore no need to invoke an association of any of the UV and the
X-ray activities with the secondary star of the \gam~ binary system. Indeed,
the variable ultraviolet diagnostics we found can be expected to be associated
with the wind or disk of the Be star, and the X-ray fluxes are in turn
correlated with them (Smith \& Robinson 1999, Robinson \& Smith 2000).

  The evidence for a correlation between X-ray and optical cycles 
continues to accumulate, as in our Fig.\,\ref{xrlcnov}. This seems to 
suggest that the X-ray production is ultimately tied to properties of
a magnetized decretion disk. This argument, along with a noncorrelation 
of epochal X-ray fluxes with respect to the 204-day binary period, led
RSH02 to suggest that a mechanism in the disk controls the conditions
for hard X-ray production. However, this does not mean that this flux
is emitted in this structure. In fact, the rapid evolution of flares is 
consistent only with densities of $\ge$ 10$^{14}$ cm$^{3}$. This fact
led SRC98 to place them on the surface of the star. The data used in
this study is adequate to show that flare aggregates cannot trigger a
response at visible wavelengths. SRC98 came to a similar conclusion
based on uncorrelated X-ray and ultraviolet continuum fluxes.

  A major unresolved issue in our understanding of the correlated X-ray
and optical cycles is the possible role of a decretion disk dynamo. If the
star's magnetic field intersects the stellar disk, it can be expected to
quench an MRI disk dynamo. Perhaps the wind's outward flow prevents the field
from crossing the disk plane. According to the geometry of strong stellar
winds in magnetic stars, this seems likely to some extent (see Smith \&
Fullerton 2004, Gagn\'e et al. 2005). At some distances from the star, the
wind finally dominates the field and opens outwards toward infinity. Thus,
the flow and the field lines no longer penetrate the disk efficiently. Under
these circumstances perhaps a dynamo can survive. The intermediate-timescale
variations are
equally puzzling.  Are they evidence for spatially local structures which
decouple from an otherwise interconnected magnetosphere/decretion
disk system?  Or, despite the similarity of their X-ray/optical scaling
with the scaling of the long cycles, is the optical flux emitted from
somewhere else, such as individual plasma clouds in the magnetosphere?

  The most important result of this paper is the discovery of a coherent 
1.2-day periodicity in both the $B$ and $V$ filters. The determination of
The determination of a coherent period clearly moves the production of the 
X-rays one step closer to the intrinsic properties of the Be star.
Like the larger optical variations of the multi-order magnetic B2p star
HD\,37776, the modulation in the \gam~ light curve 
has no accompanying $B\,-\,V$ color term. Our linking 
the \gam~  and HD\,37776 variations together implies that 
\gam~ may yet turn out to be chemically peculiar and therefore
an (albeit complex) member of the Bp class. 
However, helium is thought to remain closely coupled to hydrogen particles 
in winds of the hot stars (e.g., Hunger \& Groote 1999). 
It is not yet clear whether a mechanism exists to segregate and distribute 
helium atoms over a star as hot as $\gamma$\,Cas. 
Accordingly, it may be productive to undertake a 
search for helium line strength modulations around the 1.2-day period. 
Such a test, e.g. of He\,I $\lambda$4471 (wholly in absorption), would 
provide an important means by which to
establish whether helium rich patches associated with the early-Bp star can 
be found in stars hotter than the canonical limit (type B2) of this class.

  It is likely that the B0.5--B1 IV star HD\,110432 is a new member of the 
\gam~ ``class" (Smith \& Balona 2006).  In addition to this star, Motch et 
al. (2005) have suggested the addition of four new members to the group.
According to our picture, a \gam~ star requires a dense disk, magnetic
field (probably complex), and a rapid rotation. The rapid rotation
is necessary to insure the production of magnetic stresses between
the star and the inner region of the Be disk where the rotation law
transitions from corotation to keplerian. The high-order
field is an additional (empirical) requirement, based on the absence
of rotationally modulated UV resonance lines and H$\alpha$ emission.
These periodic emissions/absorptions over a cycle  have proved to be 
reliable spectroscopic hallmarks of dipolar magnetic Bp stars. The requirement 
of rapid rotation and a magnetic field may also mean that $\gamma$\,Cas 
stars exist near the end of their main sequence evolution stage. 
This speculation emerges from growing evidence that 
magnetic fields and CNO-processing products (including He-enhancement)
are correlated in evolved stars (Neiner et al. 2003, 
Lyubimkov et al. 2004, Huang \& Gies 2006). If additional abundance
determinations of evolved stars confirm this initial trend,
$\gamma$\,Cas analogs could be absent in very young clusters because
not enough time has elapsed to bring their fields to the surface.

We thank Mr. Lou Boyd for his continuing support of the automatic telescopes
at Fairborn Observatory and Dr. Frank Fekel for providing the Lehmann-Filhes
fit to the phase curve in Fig.~\ref{phsv29}. We also want thank Dr. David
Bohlender for pointing out the importance of HD\,37776 for the context
of this paper. It is also a pleasure to thank both Dr. Steve Cranmer and
an anonymous referee for suggestions that improved this paper.
This work was supported by NASA grant NNG05GB60C to the Catholic University 
of America and by NASA grant NCC5-511 and NSF grant HRD 97-06268 to 
Tennessee State University.

\clearpage

\centerline{\bf Figure Captions }

 \begin{figure}
\vspace*{-0.1in}
 \caption{ The m$_V$, m$_B$ scatter plot for APT observations of \gam~
during the 2000/2001 and 2001/2002 seasons. The regression lines shown 
as dashed and solid lines have slopes that differ from one another at 
a significance level of 2.4 sigma. }
\label{bvscat}
 \end{figure}

\begin{figure}
\vspace*{-0.1in}
 \caption{ The sequence of {\it RXTE/PCA} fluxes (dots) and m$_{v}$ 
magnitudes (crosses)
for \gam~ on 2004 November 9 (HJD 2453318). Error bars for the two types 
of observations are shown.  Asterisk symbols represent the X-ray data 
binned to the same sampling rate as the APT data (crosses), or about
10 minutes. The analysis of simultaneous data discussed in the text 
concerns the paired asterisks-crosses shown. }
\label{xv03}
 \end{figure}

 \begin{figure}
\vspace*{-0.1in}
 \caption{The relative $B$ and $V$ magnitude light curve on the nights of 
2003 November 19 and 20 (Julian Dates 2452962--3), shown as squares and dots
respectively (zeropoints are arbitrary).  
The dashed curves are linear regression fits to the data.
}
\label{mvdlt}
 \end{figure}

 \begin{figure}
\vspace*{-0.1in}
 \caption{The {\it RXTE}/PCA light curve for \gam~ for our program. Each dot
represents a 16-second integration. The scatter is due to rapid flaring 
and to slow chaotic undulations; the errors
in the observations, shown in the top left, are very small by comparison.
Time-zero refers to 2004 November 05.0 (i.e., HJD\,2453314.5).
The dashed line is the {\it predicted} cyclical behavior of the X-ray flux
based of the contemporaneous 85 day optical cycle determined by the robotic
Automated Photometric Telescope system located in Arizona. This curve was
computed from the X-ray/optical relation over several cycles found by
Robinson, Smith, \& Henry (2002). {\it No} time or flux adjustments have
been made in its representation.  }
\label{xrlcnov}
 \end{figure}

 \begin{figure}
\vspace*{-0.1in}
 \caption{Cycles for 1997-2004 seasons. Dashed lines denote alternative
(constant period) solutions or an undamped fit (panel g). The comb at the 
bottom of panel h shows the time interval of simultaneous APT--{\it RXTE} 
observations. The magnitude scale on each of the figures is the same 
except for season 2003/4. As in other representations, the magnitude
zero is uncertain for season 1997/8.}
\label{allcyc}
 \end{figure}

 \begin{figure}
\vspace*{-0.1in}
\caption{$Top:$  Power spectrum of the season 2 through 9 $V$-observations 
after fixing the low frequencies 0.00838, 0.01225, 0.00103, 0.00321 
day$^{-1}$.  The ordinate is the fractional reduction of the variance.
The strongest remaining frequency, marked with the large arrow, is 0.82250
$\pm$0.00001 day$^{-1}$, corresponding to a period of 1.21580 $\pm$
0.00002 days.  The smaller arrows mark the $\pm$1 day aliases of the
$\pm$0.82250 day$^{-1}$ frequency.  $Bottom:$  Power spectrum showing the 
results of fixing the four low frequencies {\it and} the 0.82250 day$^{-1}$ 
frequency.  There are no other frequencies in this range that appear 
significantly above the noise level, except for the one day aliases of 
remaining low-frequency variation.}
\label{pspect}
 \end{figure}

 \begin{figure}
\vspace*{-0.1in}
 \caption{$Top:$
Season 2 through 9 photometric V observations phased with
the 1.21851 period and the arbitrary epoch HD\,2450000. The data 
have been prewhitened to remove the low frequencies given in the text; the 
mean is also removed.  The solid line is the fit for a ``Lehmann-Filhes"
function (equation 2).  {\it Bottom}:  Data from the top panel averaged
into 100 phase bins and plotted with an expanded scale for the $y$ axis.  
Error bars give the standard deviations of the mean for each bin.  The 
solid curve is the same one in the top panel.  The binned data show good 
conformance to the now more discernible mean curve. }
\label{phsv29}
 \end{figure}

\clearpage
\begin{deluxetable}{ccccc}
\tablenum{1}
\tablewidth{0pt}
\tablecaption{PHOTOMETRIC OBSERVATIONS OF \gam}
\tablehead{

\colhead{Date} & \colhead{Var $B$} & \colhead{Var $V$}
& \colhead{Chk $B$} & \colhead{Chk $V$} \\

\colhead{(HJD $-$ 2,400,000)} & \colhead{(mag)} & \colhead{(mag)}
& \colhead{(mag)} & \colhead{(mag)}

}
\startdata
50711.7151 &   99.999 &   99.999 & $-$0.822 &   99.999 \\
50718.6966 & $-$4.393 & $-$3.641 & $-$0.818 & $-$1.216 \\
50718.9253 & $-$4.393 & $-$3.642 & $-$0.814 & $-$1.222 \\
50720.7936 & $-$4.386 & $-$3.644 & $-$0.821 & $-$1.218 \\
50720.9191 & $-$4.387 & $-$3.638 &   99.999 & $-$1.223 \\
50721.6940 &   99.999 & $-$3.648 &   99.999 &   99.999 \\
\enddata
\tablecomments{Table 1 is presented in its entirety in the electronic edition
of the {\it Astrophysical Journal} and at
http://schwab.tsuniv.edu/t3/gammacas/gammacas.html. A portion is shown here 
for guidance in data format and content.}
\end{deluxetable}

\clearpage


\begin{table}[ht!]
\tablenum{2}
\begin{center}
\caption{\label{}\centerline{Cycle Fit Parameters to Two Sinusoids\tablenotemark{*}    } }
\centerline{~}
\begin{tabular}{ccrllllc}  \hline

Year & $P_o$ & $\dot{P}/P$ & $a_1$ &  $a_2$  & $<m_v(start)>$ & $<m_v(end)>$ & 
                        $t_o$ - HJD2400000  \\
\hline 
1997/8 & 61 & 0.0  & 0.009 &  0.011  & ~~~~~~2.165  & ~~~~~~2.160 & 50750.2 \\
1998/9 & 65 & 0.0  & 0.020 & $\sim$0 & ~~~~~~2.134  & ~~~~~~2.170 & 51145.7 \\
1999/0 & 72 & -2.4$\times10^{-4}$  & 0.008 & 0.0015 & ~~~~~~2.147 & ~~~~~~2.147   & 51311.2 \\
2000/1 & 91 & 0.0  & 0.012 & 0.0012 & ~~~~~~2.143   & ~~~~~~2.143  & 51825.0 \\
2001/2 & 73  & 1.5$\times10^{-5}$ & 0.012 & 0.007 & ~~~~~~2.142 & ~~~~~~2.132     & 52123.3 \\
2002/3 & 80 & -1.0$\times10^{-5}$ & 0.0125 & 0.0125 & ~~~~~~2.1365  & ~~~~~~2.138  & 52521.8 \\
2003/4 & 80  & 0.0                 & 0.021  & 0.031  & ~~~~~~2.137   & ~~~~~~2.128  & 52805.5 \\
2004/5 & 85  & 0.0                 & 0.013  & 0.013  & ~~~~~~2.1362  & ~~~~~~2.1377 & 53197.5 \\
\hline \hline
\end{tabular}
\end{center}
\tablenotetext{}{$^*$Errors: $\delta$P = $\pm{2}$ days, Fractional $\dot{P}/P$
= ${\pm 20}$\%, $a_1$, $a_2$ = $\pm{0.001}$mags., $<m_v>$ =$\pm{0.002}$ mags.,
t$_o$ = $\pm{3}$ days. }

\end{table}

\clearpage
\begin{deluxetable}{ccccc}
\tablenum{3}
\tablewidth{0pt}
\tablecaption{Sinusoidal Fits to 1.2 day Periods in Yearly \gam~ Observations}
\tablehead{

\colhead{} & \colhead{P = 1.21581} & \colhead{P = 1.21581}
& \colhead{Control: P = 1.1976} & \colhead{Control P = 1.1976} \\

\colhead{} & \colhead{Peak-to-Peak Ampl.} & \colhead{Phase of Minimum}
& \colhead{Peak-to-Peak Ampl.} & \colhead{Phase of Minimum} \\

\colhead{Year} & \colhead{(mag)} & \colhead{(phase units)}
& \colhead{(mag)} & \colhead{(phase units)}

}
\startdata
1998/9 & 0.0066 $\pm$ 0.0015 & 0.921 $\pm$ 0.036 & 0.0085 $\pm$ 0.0015 & 0.242 $\pm$ 0.028 \\
1999/0 & 0.0066 $\pm$ 0.0015 & 0.818 $\pm$ 0.036 & 0.0109 $\pm$ 0.0015 & 0.256 $\pm$ 0.022 \\
2000/1 & 0.0028 $\pm$ 0.0018 & 0.775 $\pm$ 0.091 & 0.0061 $\pm$ 0.0017 & 0.170 $\pm$ 0.042 \\
2001/2 & 0.0095 $\pm$ 0.0014 & 0.848 $\pm$ 0.023 & 0.0071 $\pm$ 0.0014 & 0.374 $\pm$ 0.030 \\
2002/3 & 0.0053 $\pm$ 0.0017 & 0.816 $\pm$ 0.048 & 0.0075 $\pm$ 0.0016 & 0.211 $\pm$ 0.035 \\
2003/4 & 0.0040 $\pm$ 0.0012 & 0.783 $\pm$ 0.044 & 0.0039 $\pm$ 0.0012 & 0.425 $\pm$ 0.045 \\
2004/5 & 0.0112 $\pm$ 0.0009 & 0.985 $\pm$ 0.013 & 0.0051 $\pm$ 0.0010 & 0.235 $\pm$ 0.030 \\
2005/6\tablenotemark{a} & 0.0026 $\pm$ 0.0042 & 0.024 $\pm$ 0.267 & 0.0106 $\pm$ 0.0042 & 0.194 $\pm$ 0.063 \\
Sigma  &        0.0030       &       0.074       &        0.0025       &       0.089       \\
\enddata
\tablenotetext{a}{Partial season through July analyzed with only 31 observations.}
\end{deluxetable}

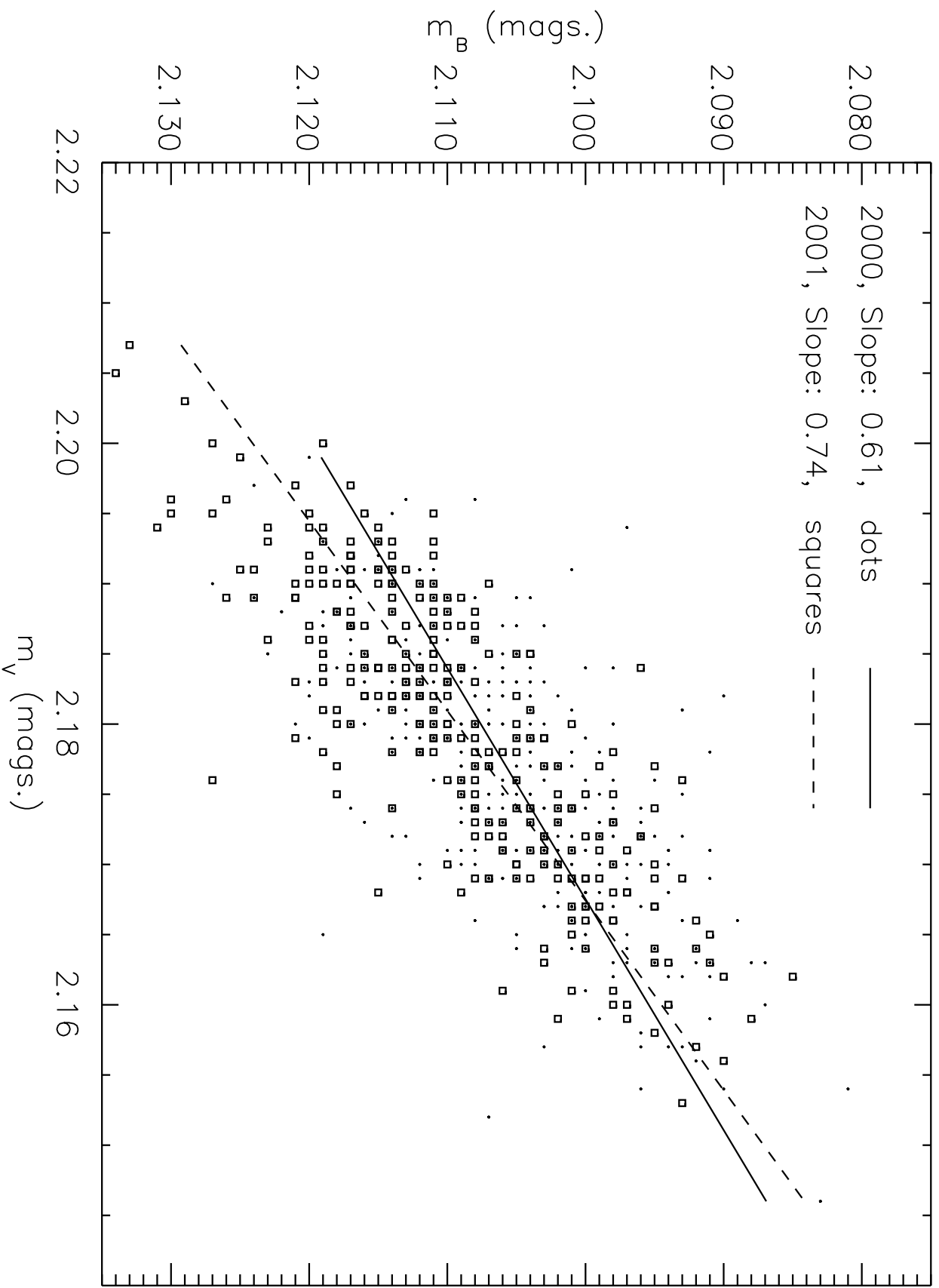
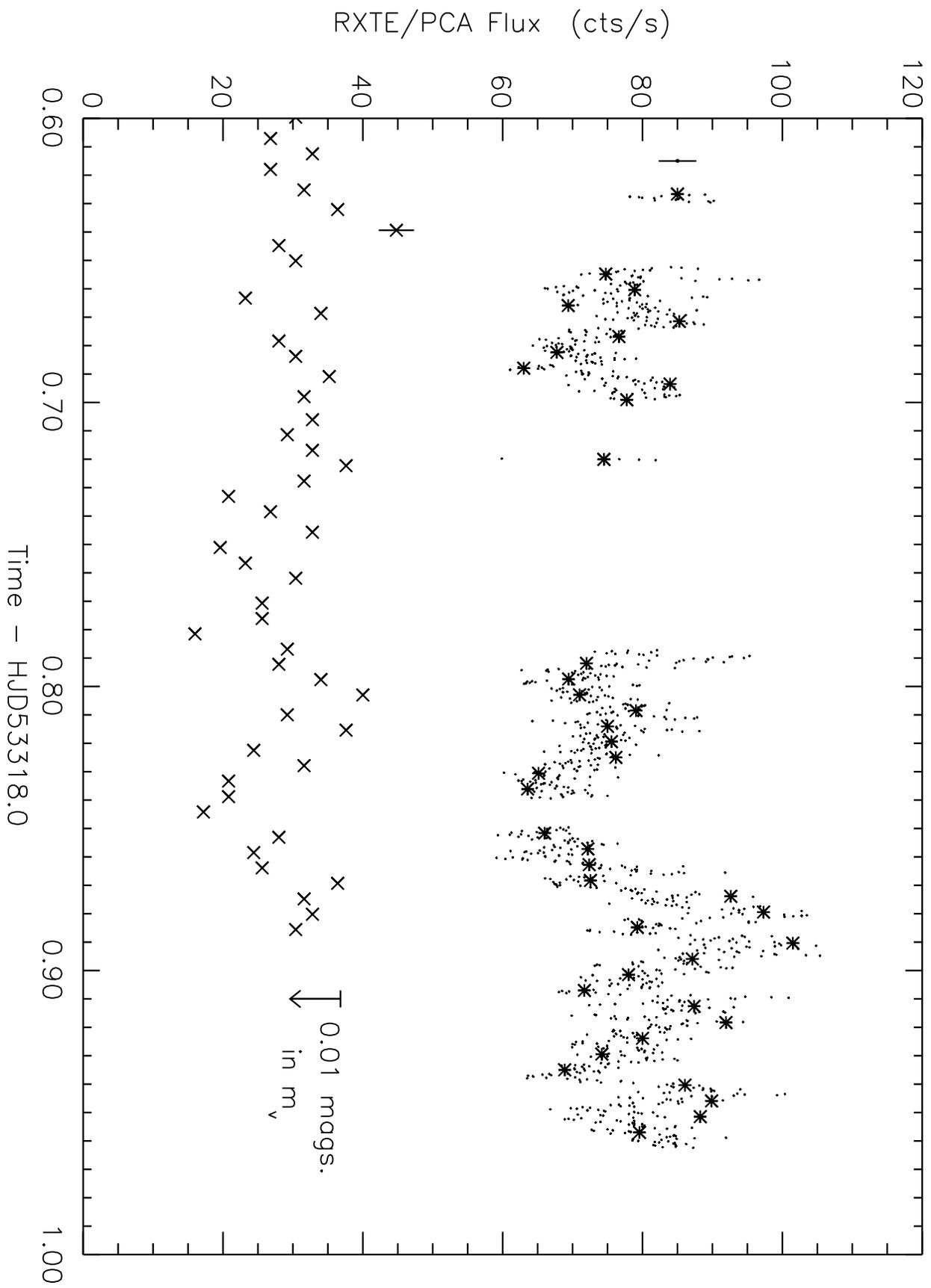
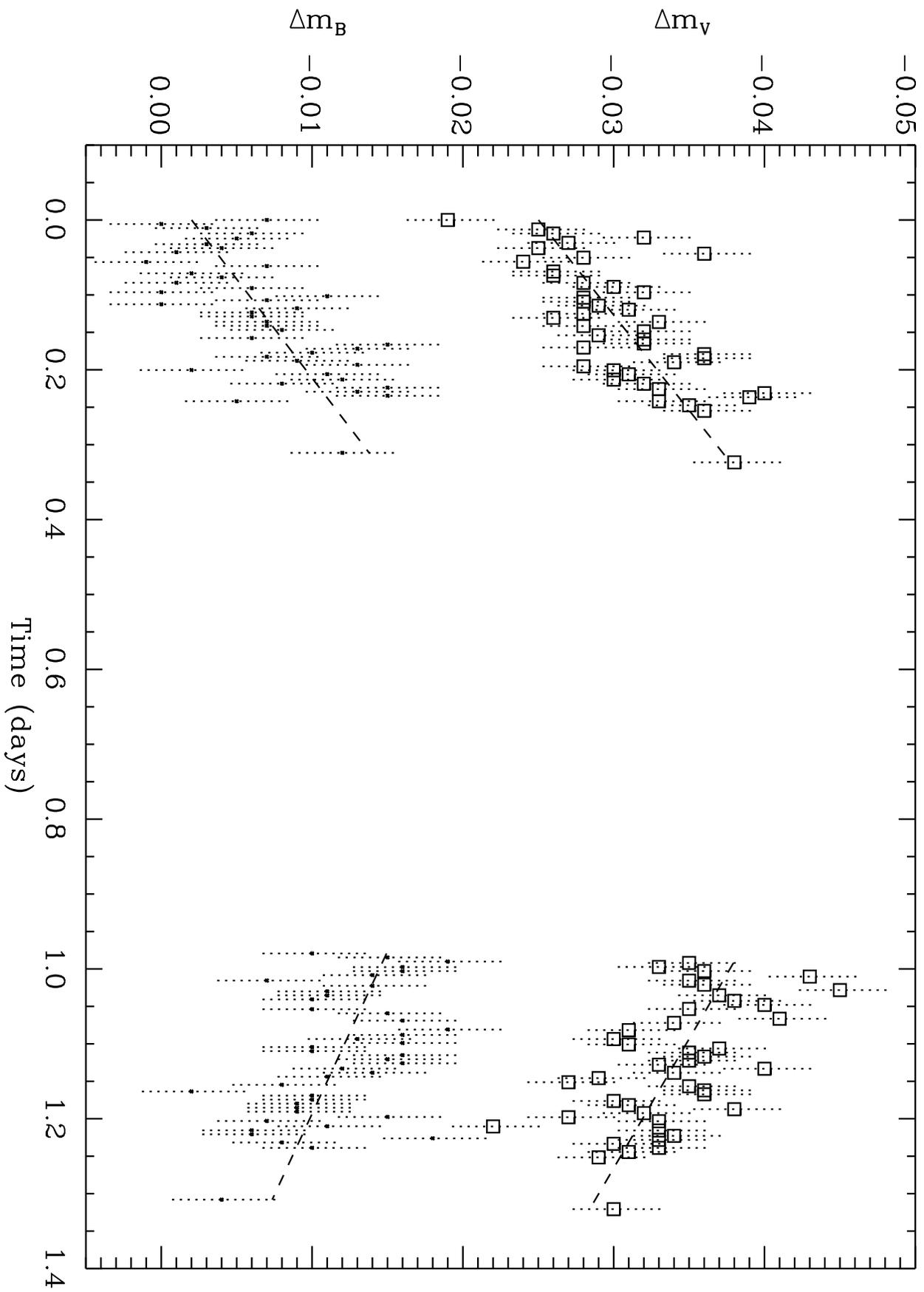
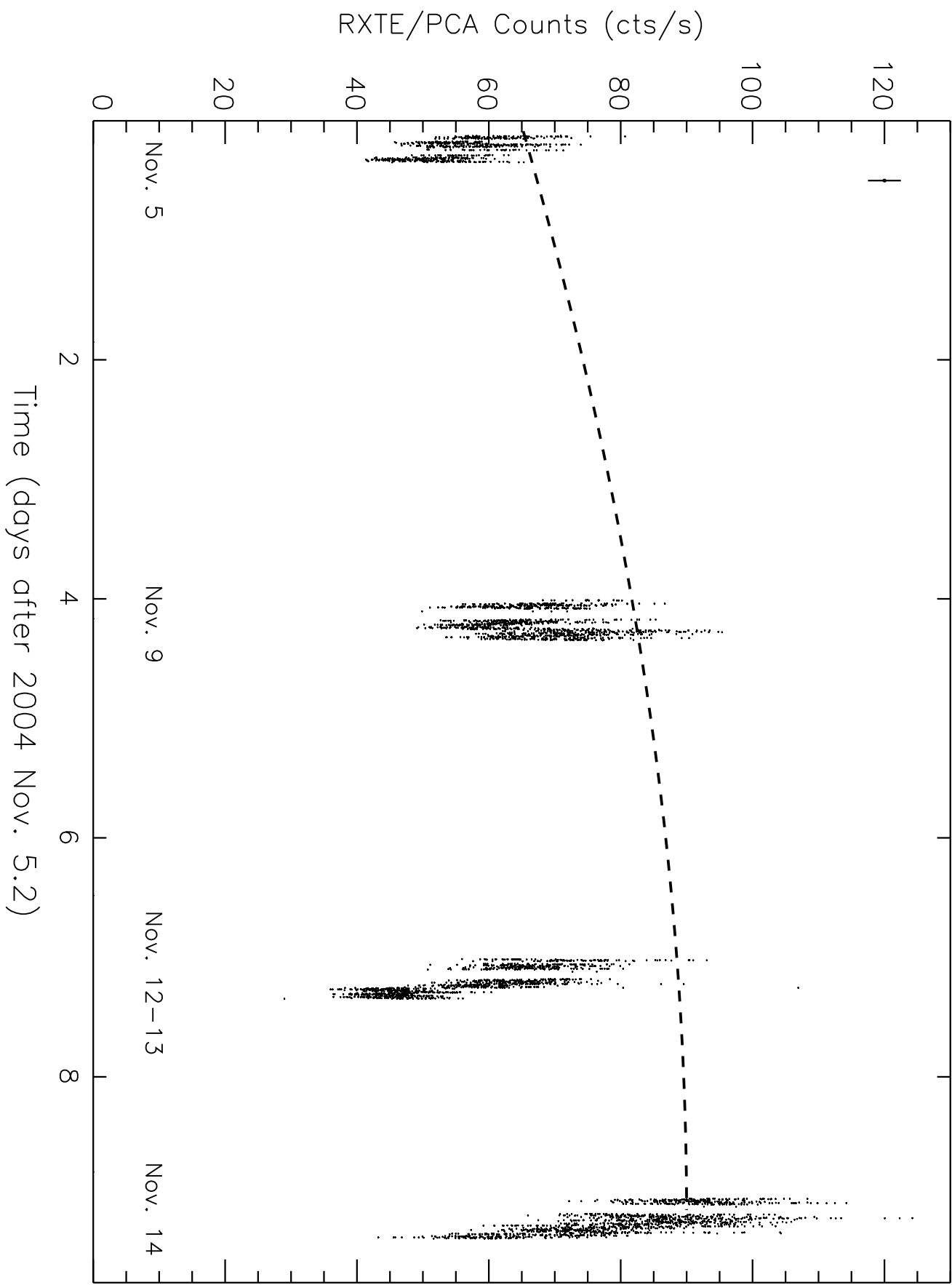
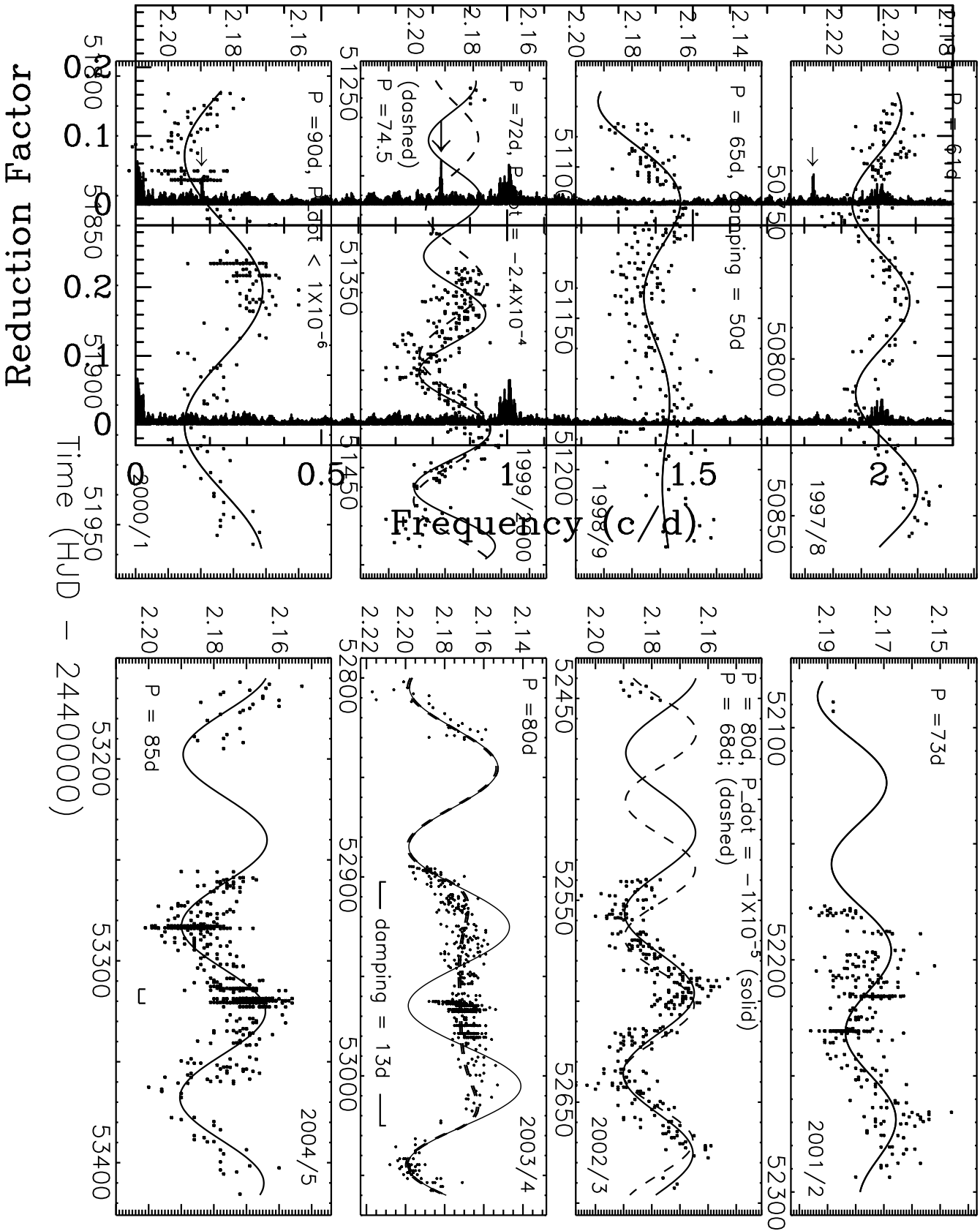
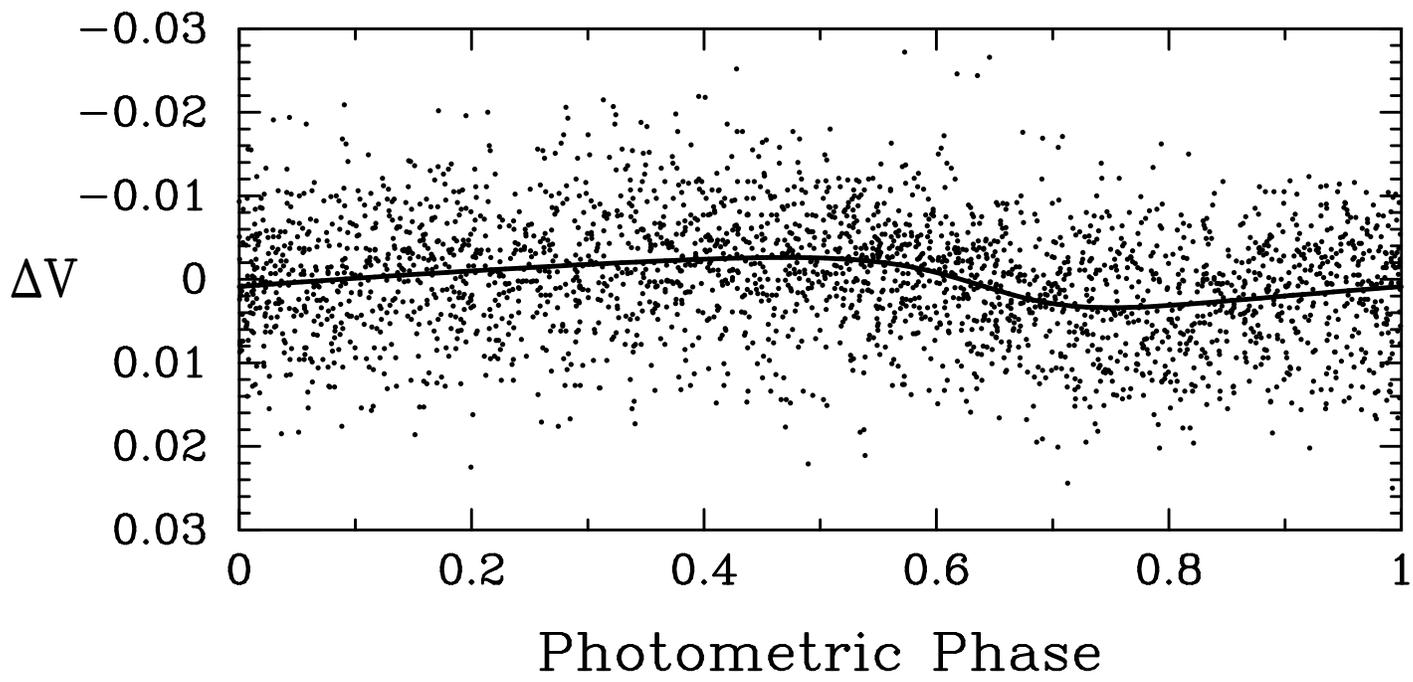
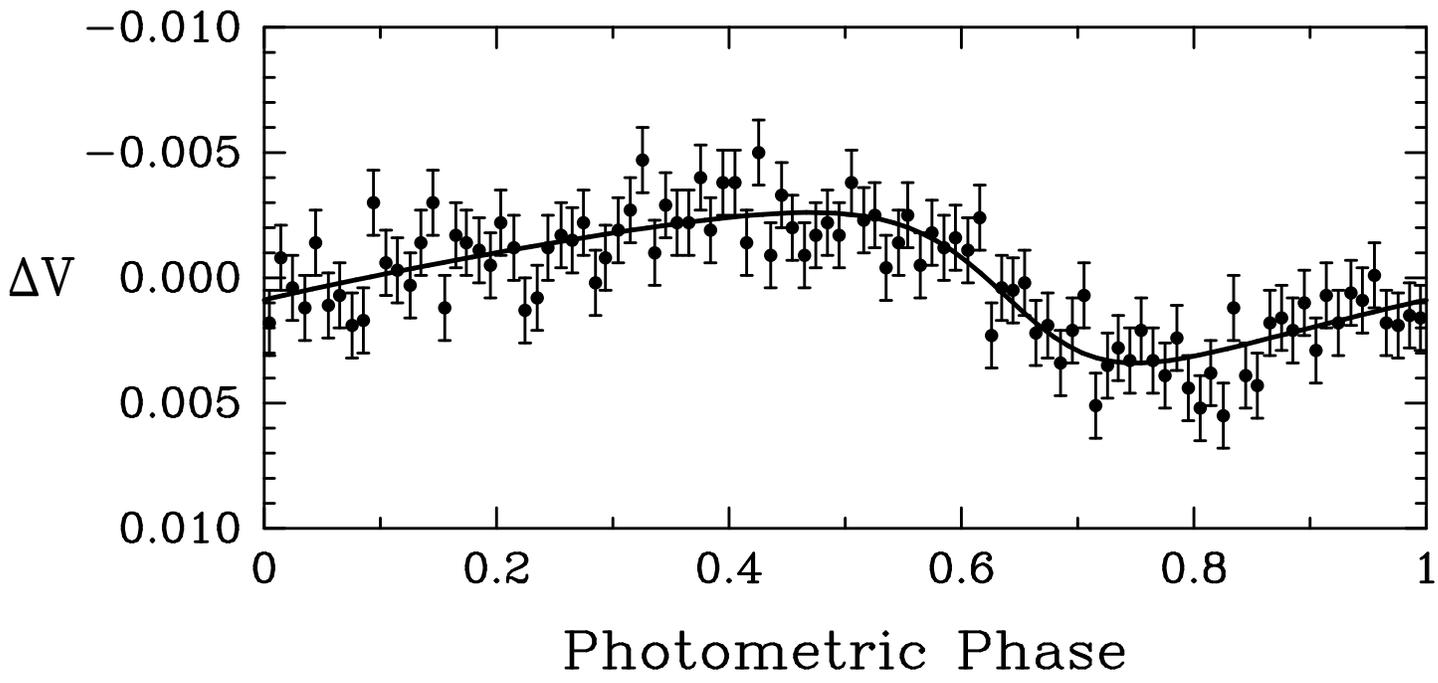
\end{document}